\documentclass[11pt]{article}
\usepackage{amsfonts, amssymb, graphicx, a4, epic, eepic, epsfig, amsmath,
setspace, lscape}
\usepackage{bbm}

\usepackage{color,soul,cancel}
\definecolor{Red}{cmyk}{0,1,1,0}
\definecolor{Blue}{cmyk}{1,1,0,0}
\definecolor{ForestGreen}{cmyk}{0.91,0,0.88,0.12}

\def\inst#1{$^{#1}$}

\def\1{\rlap{\mbox{\small\rm 1}}\kern.15em 1}

\newtheorem{theorem}{Theorem}[section]
\newtheorem{lemma}[theorem]{Lemma}
\newtheorem{proposition}[theorem]{Proposition}

\newtheorem{definition}[theorem]{Definition}
\newtheorem{corollary}[theorem]{Corollary}
\newtheorem{remark}[theorem]{Remark}

\def \H {{\mathbb H}}

\newcommand{\cX}{{\cal X}}

\def \L {{\Lambda}}

\def \m {{\mu}}

\def \s {{\sigma}}

\def \g {{\gamma}}
\def \t {{\tau}}
\def \o {{\omega}}
\def \d {{\delta}}
\def \p {{\pi}}
\def \x {{\xi}}
\def\<{{\langle}}
\def\>{{\rangle}}

\newcommand{\be}[1]{\begin{equation}\label{#1}}
\newcommand{\ee}{\end{equation}}

\newcommand{\bl}[1]{\begin{lemma}\label{#1}}
\newcommand{\el}{\end{lemma}}

\newcommand{\br}[1]{\begin{remark}\label{#1}}
\newcommand{\er}{\end{remark}}

\newcommand{\bt}[1]{\begin{theorem}\label{#1}}
\newcommand{\et}{\end{theorem}}

\newcommand{\bd}[1]{\begin{definition}\label{#1}}
\newcommand{\ed}{\end{definition}}

\newcommand{\bcl}[1]{\begin{claim}\label{#1}}
\newcommand{\ecl}{\end{claim}}

\newcommand{\bp}[1]{\begin{proposition}\label{#1}}
\newcommand{\ep}{\end{proposition}}

\newcommand{\bc}[1]{\begin{corollary}\label{#1}}
\newcommand{\ec}{\end{corollary}}

\newcommand{\bi}{\begin{itemize}}
\newcommand{\ei}{\end{itemize}}

\newcommand{\ben}{\begin{enumerate}}
\newcommand{\een}{\end{enumerate}}

\def \qed {{\square\hfill}}
\def \Z {{\mathbb Z}}

\usepackage{subfigure}
\usepackage[margin=16pt,font=small,labelfont=bf]{caption}

\usepackage{enumitem}
\usepackage{parskip}
\usepackage{amsmath}
\usepackage{xcolor}
\usepackage{mdframed}

\usepackage{tikz}
\usetikzlibrary{math}
\usetikzlibrary{intersections}
\usetikzlibrary{scopes}
\usetikzlibrary{shapes}
\usetikzlibrary{arrows,automata}
\usepackage{pgfplots}
\pgfplotsset{
        compat=1.11,
        }

\tikzstyle myBG=[line width=0.2pt,opacity=1.0]

\tikzstyle myBG=[line width=0.2pt,opacity=1.0]
\pgfkeys{/pgf/declare function={arctanh(\x) = 0.5*(ln((1+\x)/(1-\x)));}}

\begin {document}



\title{Criticality of measures on 2-d Ising configurations: from square to hexagonal graphs}

\author{
	Valentina Apollonio\inst{1}\and
	Roberto D'Autilia\inst{1}\and
	Benedetto Scoppola\inst{2} \and
	Elisabetta Scoppola\inst{1}\and
	Alessio Troiani\inst{3}}


\maketitle

\begin{center}
	{\footnotesize
		\vspace{0.3cm} \inst{1}  Dipartimento di Matematica e Fisica, Universit\`a
		Roma Tre\\
		Largo San Murialdo, 1 - 00146 Roma, Italy\\

		\vspace{0.3cm}\inst{2}  Dipartimento di Matematica, Universit\`a di Roma
		``Tor Vergata''\\
		Via della Ricerca Scientifica, 1 - 00133 Roma, Italy\\


		\vspace{0.3cm}\inst{3}  Dipartimento di Matematica ``Tullio Levi--Civita'', Universit\`a di Padova\\
		Via Trieste, 63 - 35121 Padova, Italy\\
	}

\end{center}

\begin{abstract}
\noindent
On the space of Ising configurations on the 2-d square lattice, we consider a family of non Gibbsian measures introduced by using a pair Hamiltonian, depending on an additional inertial parameter $q$. These measures are related to the usual Gibbs measure on $\Z^2$ and turn out to be the marginal of the Gibbs measure of a suitable Ising model on the hexagonal lattice.
The inertial parameter $q$ tunes the geometry of the system.
The critical behaviour and the decay of correlation functions of these measures
 are studied thanks to relation with the Random Cluster model.
\end{abstract}

\eject

\section{Introduction and definitions}
\label{Intro}
\bigskip

%
%
%
%


 Let $\L$ be a two-dimensional $2L\times 2L$ square box in $\mathbb{Z}^2$ centered at the origin and
let ${\cal B}_\L$  denote the set of all nearest neighbours
in $\L$ assuming periodic boundary conditions.
In other words ${\cal B}_\L$  consists of all pairs
$\{\langle x,y\rangle:\;x,y\in\L,\; |x-y|=1\}$,
with $|x-y|$ being the usual lattice distance in $\mathbb{Z}^2$,
plus the pairs of sites at
opposite faces of the square $\L$.
We denote by  $\cX_\L$  the set of spin configurations in $\L$, i.e., $\cX_\L = \{-1,1\}^\Lambda$.
On this spin configuration space we consider the nearest neighbours ferromagnetic Ising Hamiltonian
\be{Hs}
H(\s)=-\sum_{\langle x,y\rangle\in{\cal B}_\L}J\s_x\s_y
\ee
with $J>0$
and the  associated Gibbs measure
\be{piG}
{{\pi^G}}(\s):=\frac{1}{Z^G}e^{-H(\s)}\qquad\hbox{ with }\qquad Z^G=\sum_{\s\in \cX_\L}e^{-H(\s)}.
\ee

Looking for efficient algorithms to sample from this measure, an approximate
sampling by means of a pair Hamiltonian, adaptable to general pair interaction,
has been  introduced in \cite{dss1}. The main idea was, indeed, to define a parallel dynamics,
i.e., a Markov chain updating all spins at each time,  with
an invariant measure strictly related to $\pi^G$.
Following these ideas, a non reversible parallel dynamics with polynomial
mixing time in the size of the system has been the subject of a
successive paper \cite{dss2} where
the main ingredient  was the combination of parallel updating
and non symmetric interaction.

Define the space of pairs of configurations
$$
\cX^2_\L=\cX_\L\times\cX_\L.
$$
For each pair $(\s,\t)\in \cX^2_\L$
we define the  Hamiltonian with asymmetric interaction

\be{ham2bis3}
H(\s,\t)=
 - \sum_{x \in \L} \left[J \s_x(\t_{x^{\uparrow}} + \t_{x^{\rightarrow}}) +q\s_x \t_x\right]   = - \sum_{x \in \L}\left[J \t_x(\s_{x^{\downarrow}} + \s_{x^{\leftarrow}})  +q\t_x\s_x \right]
\ee
where $x^{\uparrow}, x^{\rightarrow}, x^{\downarrow}, x^{\leftarrow}$ are respectively the up, right, down, left neighbours of the site $x$ on the torus $(\L,{\cal B}_\L)$, $J>0$ is the ferromagnetic interaction and $q>0$ is an inertial constant.
It is straightforward to see that $H(\s,\s)=H(\s)-q|\L|$ where $H(\s)$ is the  Ising Hamiltonian given in (\ref{Hs}).
Note also that
$H(\s,\t)\not=H(\t,\s)$.


On the configuration space $\cX_\L$ we define the following family
of measures, indexed by $q$:
\be{pi}
\pi_q(\s)=\frac{1}{Z}\sum_{\t\in\cX_\L}e^{-H(\s,\t)}\qquad\hbox{ with }\qquad Z=\sum_{(\s,\t)\in \cX^2_\L}e^{-H(\s,\t)}.
\ee
These measures have been considered in the previous papers
\cite{dss1, dss2, pss1} and turn out to be the invariant
measure of the parallel dynamics  defined there.
In a more recent paper \cite{ADSST} $\pi_q(\s)$ is the
invariant measure of a reversible parallel dynamics,
the ``shaken dynamics'', modelling geological processes related
to earthquakes for a suitable choice of the parameter~$q$.

The goal of the present paper is to study,
from a static point of view,
the thermodynamical
properties and the critical behaviour of
this family of probability measures.
This analysis is performed relating $\pi_q$ to
the Gibbs measure of the Ising model
on a different lattice induced by the pair Hamiltonian
and using the standard coupling between Ising model and
Random Cluster Model (RCM).
In this context we will show that the parameter $q$ tunes the geometry of the lattice.

In the remainder of the paper, in order to lighten the notation,
we write $\pi$ in place of $\pi_q$.

The usual Gibbs measure (\ref{piG}) and the measure $\pi(\s)$ defined above are connected by the following result obtained in
\cite{dss1}, \cite{pss1} (see Theorem~1.2 in \cite{pss1}):
\bt{th0}
Define  the total variation distance, or $L_1$ distance,
between $\pi$ and $\pi^G$ as
\be{dist}
\| \pi-\pi^{G}\|_{TV}=\frac{1}{2}\sum_{\s\in \cX_{\L}}|\pi(\s)-\pi^{G}(\s)|.
\ee
Set $\d=e^{-2q}$, and let $\d$ be such that
\be{conddelta}
\lim_{|\L|\to\infty}\d^2|\L|=0,
\ee
then  there exists $\bar{J}$  such that  for any $J>\bar{J}$
\be{th01}
\lim_{|\L|\to\infty}\| \pi-\pi^G\|_{TV}=0
\ee
\et

Let us observe that the pair Hamiltonian \eqref{ham2bis3}, considering only half of the interactions (down-left), allows to interpolate between different lattices.
Indeed, as already shown in \cite{ADSST}, the space of pairs of configurations with interaction given by $H(\sigma,\tau)$ can be represented as the configuration space  $\cX_{\H}$
for the Ising model on an hexagonal lattice $\H=(V,E)$. Indeed, the hexagonal lattice $\H$ is obtained by considering two
copies $\Lambda^{1}$ and $\Lambda^{2}$ of $\Lambda$ and associating each vertex $x \in \Lambda$
to the pair $(x^{1} \in \Lambda^{1}, x^{2} \in \Lambda^{2})$. Setting {\boldmath $\sigma$}$=(\s^1,\s^2)$
with $\s^i\in \cX_{\L^i},\; i=1,2$,
and considering
the interaction defined by  $H(\sigma^{1},\sigma^{2})$ it is straightforward to observe that
$\H$ is a bipartite graph.
On this graph we distinguish two types of edges and set $E=E_J\cup E_q$.
Indeed two of the three edges exiting from each site correspond to the left and downwards interactions of strength $J$ (in the set $E_J$), while the third corresponds to the self-interaction $q$ (in the set $E_q$).\\
In other words we associate to each edge $e$ a weight
\begin{figure}
	\centering
	{

\begin{tikzpicture}[scale=0.7]

	\tikzmath{
		\xs=0.5;
		\ys=0.5;
	}
	\clip (0.1, 0.1) rectangle (9.45, 9.45);

	\draw[step=2cm,gray, help lines] (0.1,0.1) grid (9.9,9.9);
	\draw[step=2cm,gray, help lines, dashed, shift={(-\xs, -\ys)}] (0.1,0.1) grid (9.9,9.9);

	\foreach \x in {0, 2, 4, 6, 8}
		\foreach \y in {0, 2, 4, 6, 8}
			{\draw[red, thin] (\x,\y) -- (\x-\xs, \y-\ys);
			 \draw[red, thin] (\x,\y) -- (\x-\xs, \y+2-\ys);
			 \draw[red, thin] (\x,\y) -- (\x+2-\xs, \y-\ys);
			}

		\path(4,4) -- node [anchor=south east]{$q$} (4-\xs, 4-\xs);
		\path(2,4) -- node [anchor=north east, pos=0.6]{$J$} (4-\xs, 4-\ys);
		\path(4,2) -- node [anchor=north east, pos=0.6]{$J$} (4-\xs, 4-\ys);

		\path(8,8) -- node [color=black, anchor=south]{$\Lambda^1$} (10,8);
		\path[xshift=-\xs cm, yshift=-\ys cm](8,8) -- node [color=black, anchor=north]{$\Lambda^2$} (10,8);

\end{tikzpicture}}
	\caption{Interaction in the pair Hamiltonian}
	\label{fig:two_lattices}
\end{figure}
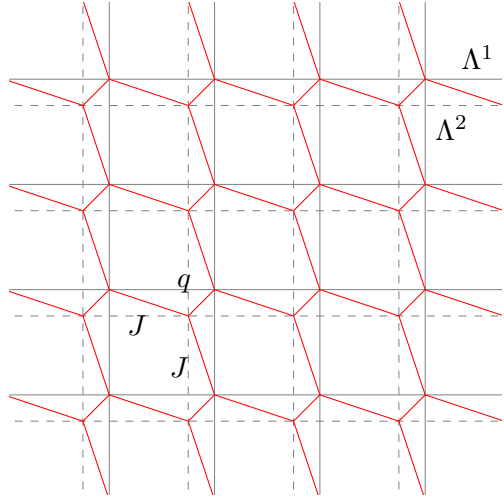
$$
{\rm J}_e= \Biggl\{
\begin{array}{lr}
J & \hbox{ if } \; e\in E_J\\
q & \hbox{ if  }\; e\in E_q
\end{array}
$$
We define the Gibbs measure for configurations {\boldmath $\sigma$}$=(\s^1,\s^2)$ on the hexagonal lattice
\be{dp2}
\pi_2(\s^1,\s^2)=\frac{e^{-H(\s^1,\s^2)}}{Z}
\ee
where the pair Hamiltonian, defined in (\ref{ham2bis3}),  has been written as
\be{ham2bis4}
H(\s^1,\s^2)=-\sum_{e\in E}{\rm J}_e\s^1_{e^1}\s^2_{e^2}
\ee
with ${e^1},{e^2}$ the two sites in $\H$ connected by the edge $e$.
Exploiting this representation, we can apply to our model
the powerful connection between Ising model and Random Cluster Model.

Assume periodic boundary conditions and define
$\Omega:=\{0,1\}^E$. For any $\o\in\Omega$ the edge $e$ is open (or present) if $\o(e)=1$.
Let $\eta(\o):=\{e\in E:\; \o(e)=1\}$ and let
$k(\o):= k(\eta (\o))$ denote the number  of connected components (or open clusters) of the graph $(V,\eta(\o))$.
Given now two parameters $p_J, p_q\in[0,1]$, by defining
$$
p_e=\Biggl\{
\begin{array}{lr}
p_J & \hbox{ if } \; e\in E_J\\
p_q & \hbox{ if  }\; e\in E_q
\end{array}
$$
we introduce the measure on $\Omega$:
\be{defPhib}
\Phi_{p_e}(\o)=\frac{1}{Z^{RC}}\Big\{\prod_{e\in E}p_e^{\o(e)}(1-p_e)^{1-\o(e)}\Big\} 2^{k(\o)}
\ee
with partition function
$$
Z^{RC}=\sum_{\o\in\Omega}\Big\{\prod_{e\in E}p_e^{\o(e)}(1-p_e)^{1-\o(e)}\Big\} 2^{k(\o)}.
$$

Following the general theory (see for instance \cite{Grbook}) we define now a coupling between our pairs of configurations {\boldmath $ \sigma$}$=(\s^1,\s^2)\in \cX_{\L}^2$ and the random cluster
configuration $\omega \in \Omega$
by the following probability mass on $\cX_{\L}^2\times\Omega$:
\be{coupling}
\m(\hbox{{\boldmath $ \sigma$}},\o)\propto \prod_{e\in E}\Big\{(1-p_e)\d_{\o(e),0}+p_e\delta_{\o(e),1}\d_e(\hbox{{\boldmath $ \sigma$}})\Big\}
\ee
where $$\d_e(\hbox{{\boldmath $ \sigma$}})=\d_{\s^1_{x}, \s^2_{y}}\quad\hbox{ for } \quad e=(x,y), \, \hbox{with} \, x \in \L^1, y \in \L^2$$

We have the following result:

\bp{p1}
If $p_J=1-e^{-2J}$ and $p_q=1-e^{-2q}$
\begin{itemize}
	\item[1)]
	the marginal on $\cX_{\L}^2$ of $\m(\hbox{{\boldmath $ \sigma$}},\o)$ is
	$$
	\m_1(\hbox{{\boldmath $ \sigma$}})=\sum_{\o\in\Omega}\m((\hbox{{\boldmath $ \sigma$}}),\o)=\pi_2(\s^1,\s^2)
	$$
	\item[2)]
	the marginal on $\Omega$ of $\m(\hbox{{\boldmath $ \sigma$}},\o)$ is
	$$
	\m_2(\o)=\sum_{\hbox{{\boldmath $ \sigma$}}\in\cX_{\L,B}^2}\m(\hbox{{\boldmath $ \sigma$}},\o)= \Phi_{p_e}(\o)
	$$
	\item[3)]
	the conditional measure on $\cX_{\L}^2$ given $\o$ is obtained by putting uniformly random spins
	on entire clusters of $\o$.
	These spins are constant on given clusters, are independent between clusters and each is uniformly distributed on the set $\{-1,+1\}$.
	\item[4)]
	the conditional measure on $\Omega$ given $\hbox{{\boldmath $ \sigma$}}$ is obtained by setting $\o(e)=0$ if \\
	$\d_e(\hbox{{\boldmath $ \sigma$}})=0$ and otherwise $\o(e)=1 $ with probability $p_J$ ($p_q$)  for $e\in E_J$ ($e\in E_q$).
	\ei
	\ep

For the proof of this proposition we refer to the clear review
by Grimmett \cite{Grbook} of the Fortuin-Kasteleyn
construction \cite{FK}, and to the rich papers \cite{ACCN} and \cite{ES} for further developments.
The coupling between these two models is robust and of wide applicability, in particular
in \cite{Grbook} the infinite-volume random-cluster measure and phase transitions are widely discussed.
With this construction we can easily prove that our model exhibits a phase transition and we can compute the strong anisotropy of the correlation functions.

Our results are presented in the next section and are proven in
Section~\eqref{sec:proofs}. In the final section
we describe some numerical aspects.
In what follows, for any $x, \, y \in V$ we will
denote by $\{x \leftrightarrow y \}$
the set of $\o\in\Omega$ for which
there exists an open path joining
the vertex $x$ with the vertex $y$.

\section{Results}
\label{Res}

The measure $\pi$, although not Gibbsian, turns out to be the marginal of the Gibbs measure
$\p_2$ of the Ising model of the hexagonal lattice
and inherits from it the thermodynamics. In other words we can extend to the non Gibbsian measure $\p$ thermodynamical relations
and the control of the critical behaviour obtained for the measure $\p_2$. To obtain these results we leverage on the well established random cluster coupling.

The first result relates the thermodynamical properties of the measures $\p$ and $\p_2$.
\bt{th1}
Consider the measure $\pi$ defined in \eqref{pi} as the marginal of the Gibbs measure on the hexagonal lattice
\be{pi2}
{{\pi_2}}(\s^1,\s^2)=\frac{1}{Z}e^{-H(\s^1,\s^2)}
\ee
with the same partition function
$$
Z=\sum_{(\s^1,\s^2)\in \cX^2_\L}e^{-H(\s^1,\s^2)}.
$$
The following relations hold:
\begin{itemize}
\item[1)] The average magnetization with respect to the measure $\pi$ and $\pi_{2}$ is the same, that is
$$
m:= \pi \Big(\frac{\sum_{x \in \L}\s_x}{|\L|}\Big) = m_2:= \pi_2 \Big(\frac{\sum_{x \in \L^1 \cup \L^2}\hbox{{\boldmath $ \sigma$}}_x}{2|\L|}\Big)
$$
\item[2)] Let $\pi^{+}$ ($\pi^{-}$) and $\pi^{+}_2$ ($\pi^{-}_2$) be the previous measures with plus (minus) boundary conditions, then for any $x \in \L$
$$
\pi^{\pm}(\s_x)= \pm \Phi_{p_e} (x^1 \leftrightarrow \partial \L^1)
$$
\item[3)] For any $x, y \in \Lambda$
$$\pi(\sigma_x\sigma_y) = \Phi_{p_e}(x^1\leftrightarrow y^1)$$
with the obvious notation $x^1, \, y^1 \in \L^1$ for the sites in the layer $\Lambda^{1}$ corresponding to vertices
$x$ and $y$ in $\Lambda$, respectively.
\end{itemize}
\et

In our second theorem we identify the critical behaviour of the system.
\begin{theorem}\label{c1}
The critical equation relating the parameters $J$ and $q$ in the measure $\p$ is given
by the equation:
\be{eq:Jc_of_q}
J_c(q)= \tanh^{-1}\Big(-\tanh q +\sqrt{\tanh ^2q+1}\Big)
\ee
\end{theorem}

\begin{remark}
 It is well known that the Gibbs measure $\pi^G$ on the square lattice exhibits
 a phase transition at $$J_c^G=\tanh^{-1}\big(\sqrt{2}-1\big)=0.4407...$$
Note that
$$
\lim_{q\to\infty}J_c(q)=J_c^G
$$
Furthermore, the curve $J_c(q)$ intersects  the line $J=q$ for
$J=\tanh^{-1}\big(\frac{\sqrt{3}}{3}\big) = 0.6585...$, corresponding to the critical value of $J$ in the
homogeneous hexagonal lattice (see Fig.~\ref{Jqplot}).
\end{remark}
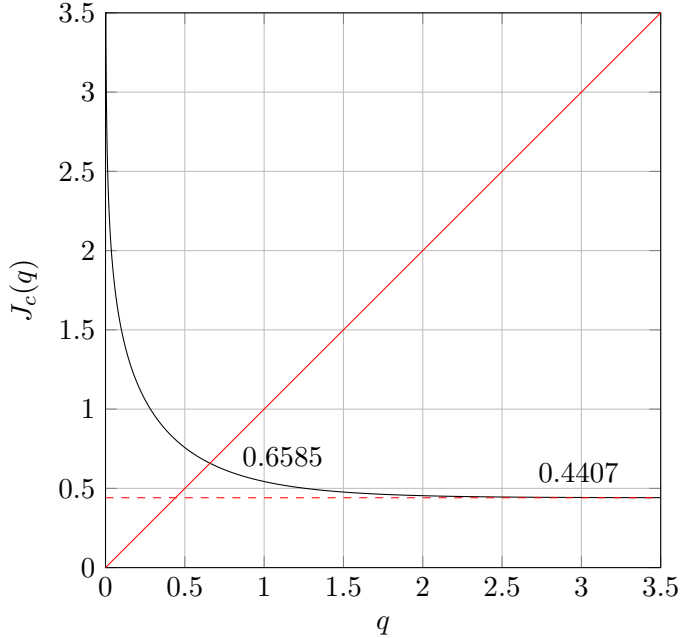
\begin{figure}[tb]
\centering
\begin{tikzpicture}
	\begin{axis}[
		x=60,
		y=60,
		xlabel={$q$},
		ylabel={$J_c(q)$},
		samples=41,
		grid,
		domain=-4:4,
		legend pos=outer north east,
		samples=1500,
		xmin=0,
		xmax=3.5,
		ymin=0,
		ymax=3.5
		]
		\draw[] (0.8,0.7) node[right,black]{$0.6585$};
		\draw[] (3.3,0.6) node[left,black]{$0.4407$};
	  \addplot [domain=0:4,mark=none,black] {arctanh(sqrt(tanh(\x)*tanh(\x)+1)-tanh(\x))};
	  \addplot [domain=0:4,mark=none,red] {x};
	  \addplot [domain=0:4,mark=none, red, dashed] {0.4406867};
	\end{axis}
\end{tikzpicture}
\caption{The function $J_c(q)$}
\label{Jqplot}
\end{figure}


The parameter $q$ tunes the geometry of the system. In fact the limit $q\to 0$ corresponds to erasing the $q$-edges obtaining, from the hexagonal lattice, independent copies of 1-$d$ Ising model. Indeed for $q\rightarrow0$ we find $J_c\rightarrow\infty$ showing the absence of phase transition for the one-dimensional Ising model. The opposite limit, $q\to\infty$, corresponds to the collapse of the hexagonal lattice into the square one, by identifying the sites connected by the $q$-edges. The case $J=q$ corresponds to the homogeneuous hexagonal graph.

The next and last result is about correlation functions and reflects the  anisotropy of the model, depending on the parameter $q$.

\bt{th3}
If the parameter $q$ is sufficiently small, for any integer $\ell \in (0, L)$  there exist two constants $c_1 < c_2$ such that
$$
\pi(\s_{(0,0)} \s_{(\ell,\ell)}) \leq c_1 < c_2 \leq \pi(\s_{(0,\ell)} \s_{(\ell,0)}).
$$
\et


\section{Proof of the results}\label{sec:proofs}

\subsection{Proof of Theorem \ref{th1}}
\begin{itemize}
\item[1)] The statement immediately follows from direct computation, indeed:
$$
m= \sum_{\substack{\s}}\frac{\sum_{\substack{x \in \L}} \s_x}{|\L|}\cdot \sum_{\substack{\t}} \frac{e^{-H(\s,\t)}}{Z}= \frac{1}{2}\sum_{(\s,\t)}\frac{\sum_{x \in \L} (\s_x+\t_x)}{|\L|} \cdot \frac{e^{-H(\s,\t)}}{Z}= m_2
$$
where the second equality follows by a symmetry argument.

\item[2)]
The standard coupling between Ising and the RCM on $\H$ yields
\begin{align*}
\pi^{+}(\s_x) & =\sum_\s  \s_x \pi^{+}(\s)= \sum_{\hbox{{\boldmath $ \sigma$}}} \s^1_{x^1} \pi^+_2(\hbox{{\boldmath $ \sigma$}})= \pi^+_2(\s^1_{x^1}) \\
& =  \sum_{\o \in \Omega} \sum_{\hbox{{\boldmath $ \sigma$}}} \mu({\hbox{{\boldmath $ \sigma$}}}, \o) \s^1_{x^1} \big(\mathbbm{1}_{x^1 \leftrightarrow \partial \Lambda^1}+\mathbbm{1}_{x^1 \nleftrightarrow \partial \Lambda^1}\big)\\
& =\Phi_{p_e}(x^1 \leftrightarrow \partial \Lambda^1) +\sum_{\o \in \Omega} \sum_{\hbox{{\boldmath $ \sigma$}}} \Big[ \mu({\hbox{{\boldmath $ \sigma$}}}, \o |\o)\s^1_{x^1} \mathbbm{1}_{x^1 \nleftrightarrow \partial \Lambda^1} \Big] \Phi_{p_e}(\o)  \\
&  =\Phi_{p_e}(x^1 \leftrightarrow \partial \Lambda^1)
\end{align*}
since by proposition \ref{p1} the square bracket vanishes. The minus boundary conditions can be treated in the same way.

\item[3)] The proof of point (3) can be obtained following the same argument.

\hfill{$\qed$}
\end{itemize}


\subsection{Proof of Theorem \ref{c1}}

As shown in \cite{CDC}, for a planar weighted graph $G= (V, E)$ that is non degenerate,
finite and doubly periodic, the critical curve of the Hamiltonian
$$
H(\s)= - \sum_{e=\{u, v\} \in E} {\rm J}_e \s_u \s_v
$$
is the unique solution of the equation
\begin{equation}
	\sum_{\gamma\in{\mathcal E}_0(G)}\prod_{e\in\gamma}\tanh{ {\rm J}_e}=\sum_{\gamma\in{\mathcal E}_1(G)}\prod_{e\in\gamma}\tanh { {\rm J}_e}
\label{eq_crit}
\end{equation}

where ${\mathcal E (G)}$ is the set of the even subgraphs of $G$, i.e., the set of subgraphs $\g$ of $G$ such that each vertex of $G$ is an endvertex of an even number of edges of $\g$, ${\mathcal E}_0(G)$ is the set of the even subgraphs of $G$ winding an even number of times around each of the two dimensions of the torus and ${\mathcal E}_1(G)={\mathcal E}(G)\setminus{\mathcal E}_0(G)$.
The main step in the proof of this result is to show that the free energy per fundamental domain can be expressed in terms of the Kac-Ward determinants.

In our case the hexagonal lattice $\H=(V,E)$  satisfies the conditions of the theorem 1.1 in \cite{CDC} 
and the equation (\ref{eq_crit}) can be obtained by periodically glueing on the torus the cell represented in Fig.~\ref{cella_esagono}.\\
\begin{figure}[tbh]
\centering
\begin{tikzpicture}
    \draw[gray!10, fill=gray!10, ] (-2,0) -- (0,1) -- (2,0) -- (0,-1) -- cycle;
	\draw (-0.5,0)--(0.5,0);
	\draw (0.5,0)--(1,1/2);
	\draw (0.5,0)--(1,-1/2);
	\draw (-0.5,0)--(-1,1/2);
	\draw (-0.5,0)--(-1,-1/2);
	\filldraw [gray] (-0.5,0) circle (2pt) (0.5,0) circle (2pt);
	\node at (-1,0) {$J$};
	\node at (1,0) {$J$};
	\node at (0,0.2) {$q$};
\begin{scope}[shift={(-4.1,-2.1)}]
    \draw[gray!10, fill=gray!10, ] (-2,0) -- (0,1) -- (2,0) -- (0,-1) -- cycle;
	\draw (-0.5,0)--(0.5,0);
	\draw (0.5,0)--(1,1/2);
	\draw (-0.5,0)--(-1,-1/2);
	\filldraw [gray] (-0.5,0) circle (2pt) (0.5,0) circle (2pt);
	\node at (1,0) {$J$};
	\node at (0,0.2) {$q$};
\end{scope}
\begin{scope}[shift={(0,-2.1)}]
    \draw[gray!10, fill=gray!10, ] (-2,0) -- (0,1) -- (2,0) -- (0,-1) -- cycle;
	\draw (-0.5,0)--(0.5,0);
	\draw (0.5,0)--(1,-1/2);
	\draw (-0.5,0)--(-1,1/2);
	\filldraw [gray] (-0.5,0) circle (2pt) (0.5,0) circle (2pt);
	\node at (-1,0) {$J$};
	\node at (0,0.2) {$q$};
\end{scope}
\begin{scope}[shift={(4.1,-2.1)}]
    \draw[gray!10, fill=gray!10, ] (-2,0) -- (0,1) -- (2,0) -- (0,-1) -- cycle;
	\draw (0.5,0)--(1,1/2);
	\draw (0.5,0)--(1,-1/2);
	\draw (-0.5,0)--(-1,1/2);
	\draw (-0.5,0)--(-1,-1/2);
	\filldraw [gray] (-0.5,0) circle (2pt) (0.5,0) circle (2pt);
	\node at (-1,0) {$J$};
	\node at (1,0) {$J$};
\end{scope}
\end{tikzpicture}
\caption{The elementary cell on the torus and the three corresponding even subgraphs $\gamma\in{\mathcal E}_1$}
\label{cella_esagono}
\end{figure}
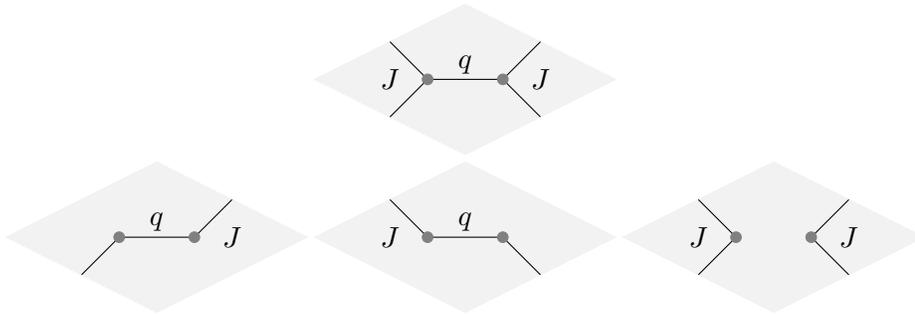\ \\\\
The explicit form of the equation is
\begin{equation}\label{eqcritJq}
1=2\tanh J\tanh q+\tanh^2 J
\end{equation}
where on the r.h.s. we have the sum of the contributions from the three even subgraphs in ${\mathcal E}_1$  shown in Fig.~\ref{cella_esagono} while the $1$ on the l.h.s. is the contribution of the unique graph in ${\mathcal E}_0$ without edges.

Solving equation \eqref{eqcritJq} w.r.t. $J$ gives the curve
\begin{equation}
	J(q)=\tanh^{-1}\big(\sqrt{\tanh^2q+1}-\tanh q\big)
\label{J_di_q}
\end{equation}
represented in Fig.~\ref{Jqplot}


\hfill{$\qed$}

\subsection{Proof of Theorem \ref{th3}}


Let $\gamma \subset E$ be a path of open edges connecting two vertices $x^{1}$, $y^{1}$ $\in \L^{1}$. 
We introduce the notation $\eta(\omega) \supset \gamma$ to identify all the configurations $\o \in \Omega$ such that $\omega(e)=1, \, \forall e \in \gamma $. By definition
\begin{align*}
\Phi_{p_e}( x^{1} \leftrightarrow y^{1}) & = \sum_{\gamma: x^{1} \leftrightarrow y^{1} } \;\,  \sum_{\substack{\omega \in \Omega \, :\\ \, \eta(\omega) \supset \gamma}} \Phi_{p_e}(\o) \\ & = \frac{1}{Z^{RC}} \sum_{\gamma: x^{1} \leftrightarrow y^{1}} \Biggl(\prod_{e\in \gamma} p_e\Biggr) \sum_{\o' \in \{0,1\}^{E \setminus \gamma}} \Biggl(\prod_{e \in E \setminus \gamma} p_e^{\o'(e)} (1-p_e)^{1-\o'(e)}\Biggr) 2^{k(\eta(\o')\cup \gamma)}  \\ &
= \frac{1}{Z^{RC}} \sum_{\gamma: x^{1} \leftrightarrow y^{1}}\Biggl(\prod_{e\in \gamma} p_e\Biggr) Z_{\gamma}
\end{align*}
where
$$
Z_{\gamma}=\sum_{\o' \in \{0,1\}^{E \setminus \gamma}} \Biggl(\prod_{e \in E \setminus \gamma} p_e^{\o'(e)} (1-p_e)^{1-\o'(e)}\Biggr) 2^{k(\eta(\o')\cup \gamma)}
$$
\textbf{Upper bound}

For any path $\gamma$ and any configuration $\o \in \Omega$ we denote by $\o'$ the restriction of $\o$ to the set of edges in ${E \setminus \gamma}$ and by $\o''$ its restriction to the set of edges in ${ \gamma}$ .

Since $k(\eta(\o))\geq k(\eta(\o') \cup \g)$ we can state the following inequality for the partition function
\begin{align*}
Z^{RC} & \geq \sum_{\o \in \Omega}\Biggl(\prod_{e\in \gamma} p_e^{\o(e)} (1-p_e)^{1-\o(e)}\Biggr)\Biggl(\prod_{e\in E \setminus \gamma} p_e^{\o(e)} (1-p_e)^{1-\o(e)}\Biggr) 2^{k(\eta(\o') \cup \g )} \\
& =  \Biggl(\sum_{\o'' \in \{0,1\}^{|\g|}} \prod_{e\in \gamma} p_e^{\o''(e)} (1-p_e)^{1-\o''(e)}\Biggr) Z_{\g}= Z_{\g}
\end{align*}
This observation implies
\be{corr}
\Phi_{p_e}(x^{1} \leftrightarrow y^{1}) \leq \sum_{\gamma: x^{1} \leftrightarrow y^{1}}\Bigl(\prod_{e\in \gamma} p_e\Bigr)
\ee


\begin{figure}[h]
\centering
\subfigure[]
{
\begin{tikzpicture}[scale=.7]

	\tikzmath{
		\xs=0.5;
		\ys=0.5;
	}

	\clip (0.1, 0.1) rectangle (9.45, 9.45);

	\draw[step=2,gray, help lines] (0.1,0.1) grid (9.9,9.9);

	\foreach \x in {0, 2, 4, 6, 8}
		\foreach \y in {0, 2, 4, 6, 8}
			{\draw[red, very thin] (\x,\y) -- (\x-\xs, \y-\ys);
			 \draw[black, very thin] (\x,\y) -- (\x-\xs, \y+2-\ys);
			 \draw[black, very thin] (\x,\y) -- (\x+2-\xs, \y-\ys);
			}
	\foreach \x in {2, 3, 4, 5, 6, 7, 8}
		{\draw[help lines, dashed] (\x-\xs/2, \x-\xs/2) -- +(-45:10);
		 \draw[help lines, dashed] (\x-\xs/2, \x-\xs/2) -- +(135:10);
		}

	\draw (2,2) node [color=black, anchor= south west] {$(0,0)$};
	\draw (8,8) node [color=black, anchor= south west] {$(\ell,\ell)$};
	\draw (8,2) node [color=black, anchor= south west] {$(\ell,0)$};
	\draw (2,8) node [color=black, anchor= south west] {$(0,\ell)$};

	\draw[blue, very thick] (2,2) -- (2-\xs, 4-\xs) -- (2,4) -- (2-\xs, 6-\xs) --
						(2,6) -- (4-\xs, 6-\xs) -- (4,4) -- (4-\xs, 4-\xs) --
						(4,2) -- (6-\xs, 2-\xs) -- (6,2) -- (6-\xs, 4-\xs)--
						(6,4) -- (6-\xs, 6-\xs) -- (6,6) -- (6-\xs, 8-\xs) --
						(6,8) -- (8-\xs, 8-\xs) -- (8,8);

%

\end{tikzpicture}}
\hfill
\subfigure[]
{
\begin{tikzpicture}[scale=.7]

	\tikzmath{
		\xs=0.5;
		\ys=0.5;
	}

	\clip (0.1, 0.1) rectangle (9.45, 9.45);

	\draw[step=2,gray, help lines] (0.1,0.1) grid (9.9,9.9);

	\foreach \x in {0, 2, 4, 6, 8}
		\foreach \y in {0, 2, 4, 6, 8}
			{\draw[red, very thin] (\x,\y) -- (\x-\xs, \y-\ys);
			 \draw[black, very thin] (\x,\y) -- (\x-\xs, \y+2-\ys);
			 \draw[black, very thin] (\x,\y) -- (\x+2-\xs, \y-\ys);
			}
	\foreach \x in {2, 3, 4, 5, 6, 7, 8}
		{\draw[help lines, dashed] (\x-\xs/2, \x-\xs/2) -- +(-45:10);
		 \draw[help lines, dashed] (\x-\xs/2, \x-\xs/2) -- +(135:10);
		}

	\draw (2,2) node [color=black, anchor= south west] {$(0,0)$};
	\draw (8,8) node [color=black, anchor= south west] {$(\ell,\ell)$};
	\draw (8,2) node [color=black, anchor= south west] {$(\ell,0)$};
	\draw (2,8) node [color=black, anchor= south west] {$(0,\ell)$};

	\draw[blue, very thick] (2,8) -- (4-\xs, 8-\xs) -- (4,6) -- (6-\xs, 6-\xs) --
						(6,4) -- (8-\xs, 4-\xs) -- (8,2);

%

\end{tikzpicture}}
\caption{Lattice $\H$ with the slices used for the estimation of the correlation functions. Picture (a) shows an example of path
$\gamma: (0,0) \leftrightarrow (\ell, \ell)$. Picture (b) shows the diagonal path
$\gamma^\star$}
\label{fig:slicedHexLattice}
\end{figure}
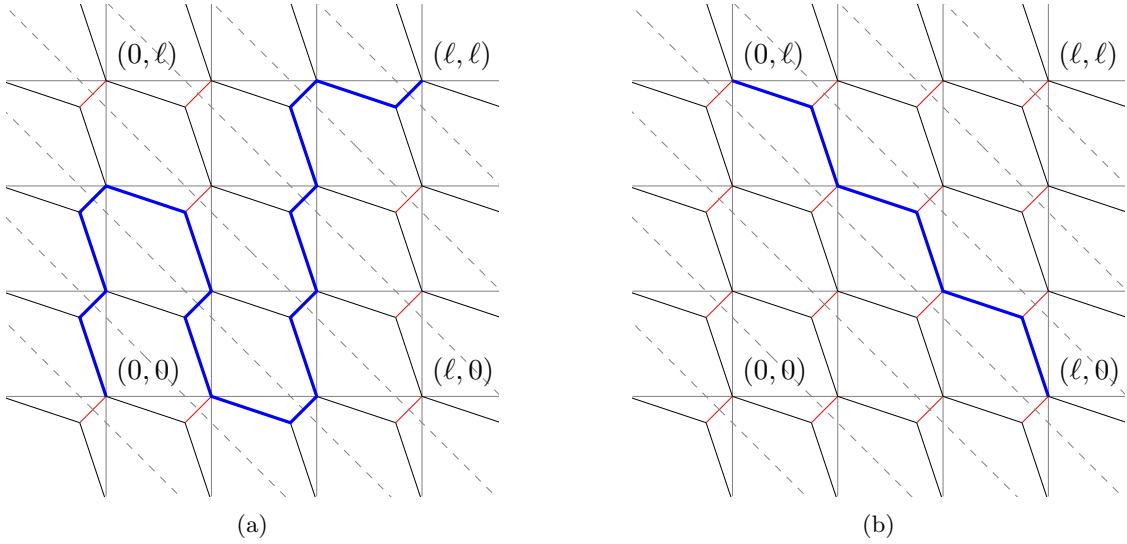

Now let us suppose to slice the lattice $\H$ as in Fig.~\ref{fig:slicedHexLattice}. It is easy to see that each path $\gamma: (0,0) \leftrightarrow (\ell, \ell)$ must visit all slices separating $(0,0)$ and $(\ell, \ell)$ and therefore it crosses at least $2 \ell$ $q$-edges
(see Fig.~\ref{fig:slicedHexLattice}(a)).
 We give an upper bound for the sum in \eqref{corr} in terms of possible crossing-paths that start in $(0,0)$ and stop in the slice which contains $(\ell, \ell)$.
The transition from one slice to the other is determined by the crossing of a $q$-edge. After a $q$-edge has been crossed the path must traverse an arbitrary number of $J$-edges, either on the left or on the right, before crossing the next $q$-edge.
Denoting by $\Gamma(n, 2\ell)$ the number of one dimensional random walks between slices of lenght $n$ arriving at distance $2\ell$ from the origin, we can write
\begin{align*}
\Phi_{p_e}((0,0)\leftrightarrow (\ell, \ell)) & \leq \sum_{n=2\ell}^{\infty} \Gamma(n, 2\ell) (2p_q p_J)^n \Biggl(\sum_{m=0}^{\infty}p_J^{m} \Biggr)^n =
\sum_{n=2\ell}^{\infty} \Gamma(n, 2\ell)(2p_qp_J)^n  \Biggl(\frac{1}{1-p_J}\Biggr)^n \\
& =  \sum_{n=2 \ell}^{\infty} \binom{n}{\frac{n+2\ell}{2}} \Biggl(\frac{2p_qp_J}{1-p_J}\Biggr)^n \leq
\sum_{n=2 \ell}^{\infty} \Biggl(\frac{4p_qp_J}{1-p_J}\Biggr)^n.
\end{align*}
The last sum converges if $q$ is sufficiently small so that the parameters $p_J$ and $p_q$ satisfy the  condition
$
\frac{4p_qp_J}{1-p_J} < 1
$
and we get
$$
c_1=\frac{\left(\frac{4p_qp_J}{1-p_J}\right)^{2\ell}}{1-\left(\frac{4p_qp_J}{1-p_J}\right)}\cdot
$$

\textbf{Lower bound}

We introduce the diagonal path $\gamma^*$ connecting $(0,\ell)$ and $(\ell, 0)$ remaining in the same slice
as in Fig.~\ref{fig:slicedHexLattice}(b) and $\bar{\g}= \gamma^*\cup \partial \g^*$.
	Let $Z_{E\setminus \bar{\g}}$ be the partition function of the Random Cluster Model defined on the graph $\H_{\bar{\g}}= (V,E\setminus\bar{\g})$.
	By Theorem (3.60) in \cite{Grbook} we have that
	$Z_{E\setminus \bar{\g}} \ge Z_{E} = Z^{RC}$ and hence we can give
	a lower bound for the correlation function as follows
\begin{align*}
& \Phi_{p_e}((0,\ell) \leftrightarrow (\ell, 0)) \geq 
\\
& \geq \frac{1}{Z^{RC}} \left(\prod_{e\in \g^*}p_e \right)
	\sum_{\o'' \in \{0,1\}^{E \setminus \bar{\g}}}\left[\prod_{e \in \partial \g^*}(1-p_e)\right]\left[\prod_{e \in E \setminus \bar{\g}} p_e^{\o''(e)}(1-p_e)^{1-\o''(e)}\right] 2^{k(\o'')+1}\\
&  =  \frac{1}{Z^{RC}} \left(\prod_{e\in \g^*}p_e \right) \left[\prod_{e \in \partial \g^*}(1-p_e)\right] 2 Z_{E\setminus \bar{\g}} \\
& \geq 2 \left(\prod_{e\in \g^*}p_e \right) \left[\prod_{e \in \partial \g^*}(1-p_e)\right]\\
&  = 2e^{-4J}(1-e^{-2J})^{2 \ell}e^{-2q(2 \ell +1)}=c_2.
\end{align*}

If $q$ is sufficiently small such that, for instance,
\begin{align*}
	\frac{4p_q}{(1-p_q)(1-p_J)} < \frac{1}{2}
\end{align*}
and
\begin{align*}
	\left(\frac{p_q}{1-p_q}\right)^2 < \frac{(1-p_J)^3(1-p_q)}{16},
\end{align*}
we immediately get $c_1 < c_2$. Note that the first of these two conditions is stronger than
$ \frac{4p_qp_J}{1-p_J} < 1 $ and, therefore, $c_1$ is well defined.
\hfill{$\qed$}

\section{Numerical indications}
The measure $\pi$ is not a Gibbs
measure. However it is possible to sample
from it in an effective
way by drawing samples from the Gibbs
measure $\pi_2$. Remarkably, this sampling can be performed
in reasonably short times even for values
of the parameters close to
the critical ones.


%

To this purpose, consider the ``shaken dynamics'' introduced in \cite{ADSST}.
This dynamics can be seen as a dynamics
on $\cX_{\H}$ that, alternatively,
updates the spins in $\Lambda^{1}$ and in $\Lambda^{2}$.
In \cite{ADSST}, Theorem~2.2, it has been shown that the equilibrium measure of the
shaken dynamics, regarded as a dynamics on $\cX_{\H}$, is the Gibbs measure $\pi_2$.

This parallel dynamics preserves
the natural partial ordering between
Ising configurations.
Consequently, it allows to effectively exploiting massively parallel computing
to draw unbiased samples from $\pi_2$
using perfect sampling techniques \cite{PW, WIL}. To draw a sample from $\pi$ it is, therefore, enough
to draw a configuration from $\pi_2$ and look at the sub-configuration on the layer $\Lambda^1$.
A more detailed numerical analysis of the shaken dynamics will be the topic of a forthcoming
paper \cite{ANT}.

\begin{figure}[h!]
\centering
\hfill
\subfigure[]
{\includegraphics[width=0.4\textwidth]{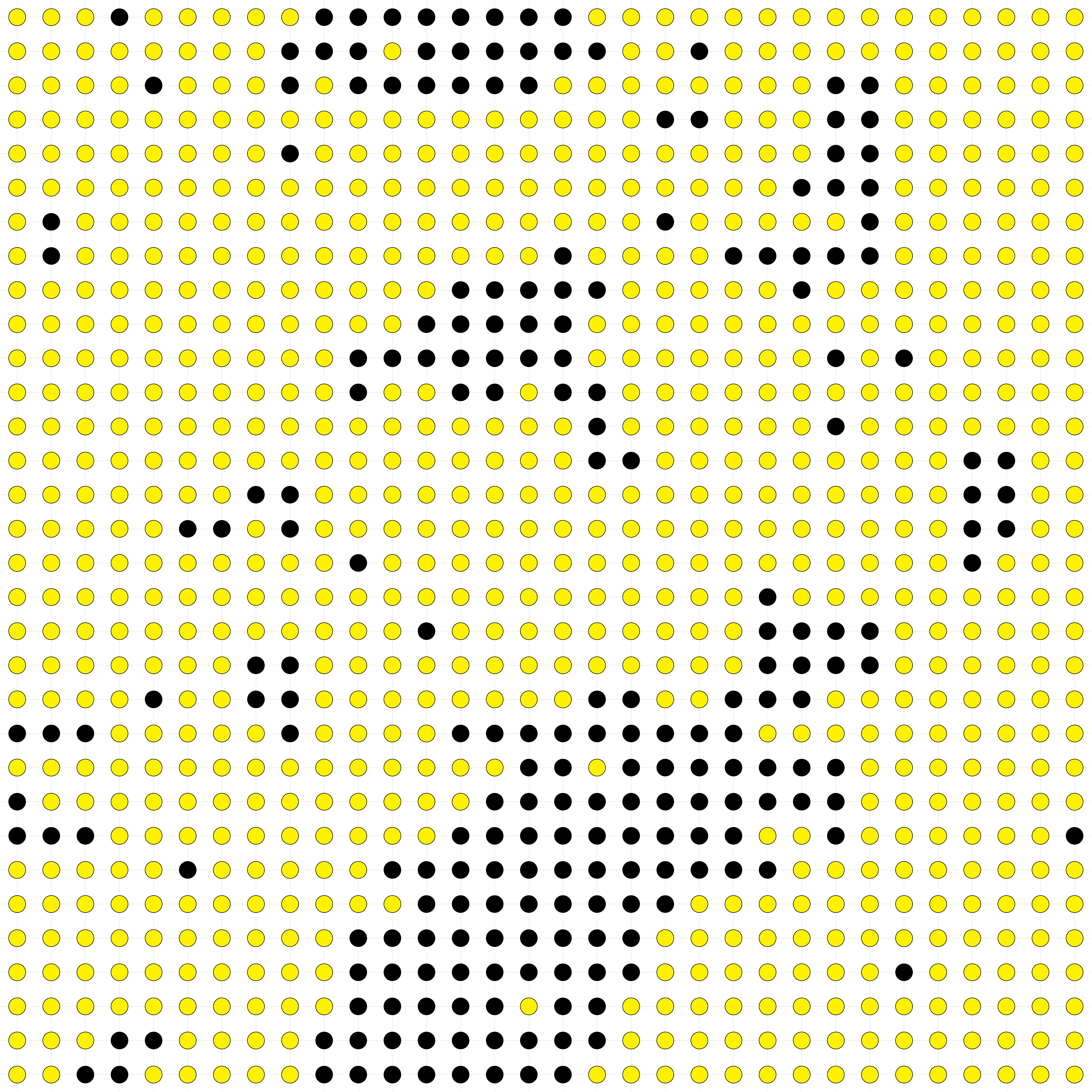}}
\hfill
\subfigure[]
{\includegraphics[width=0.4\textwidth]{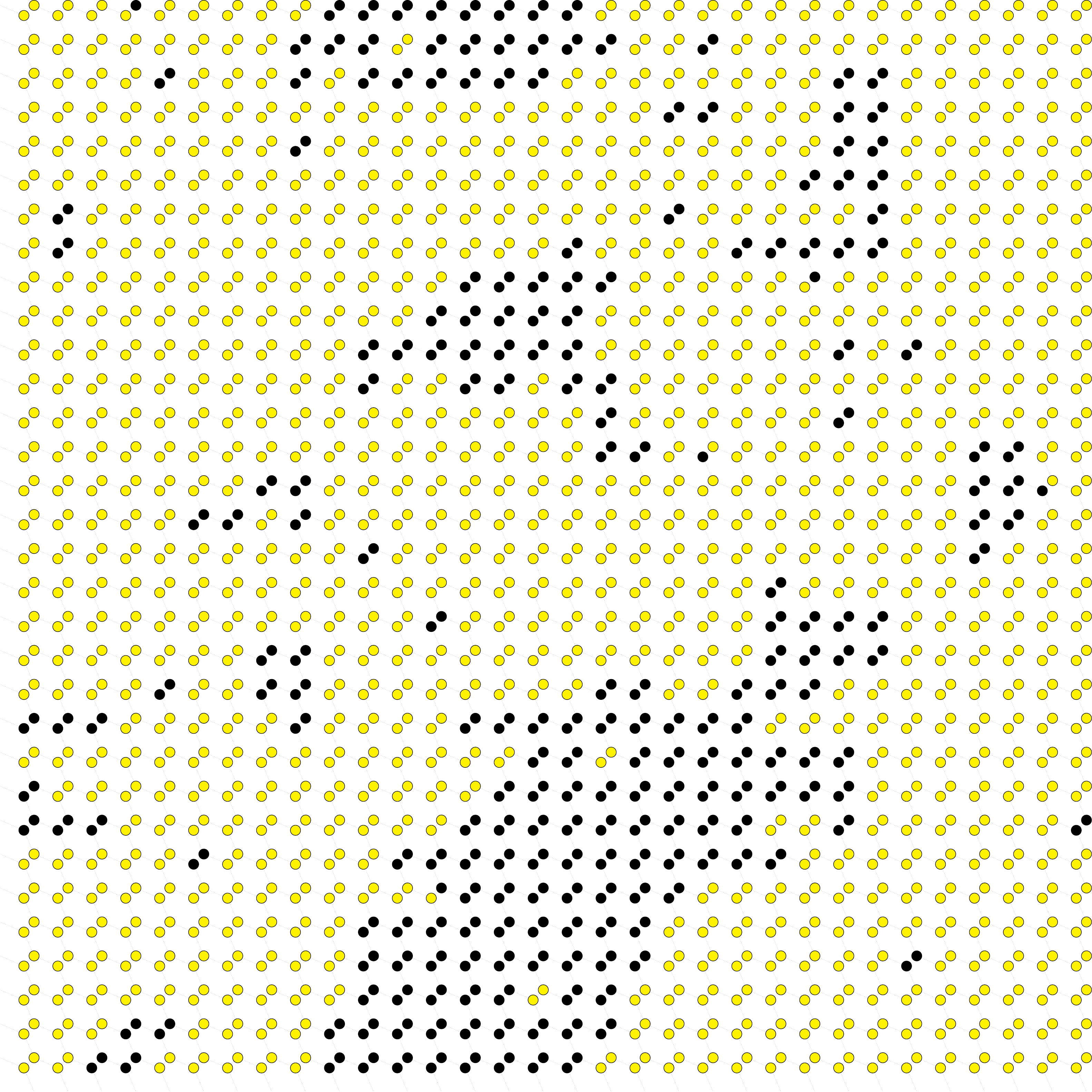}}
\hfill
\caption{$J = 0.44$, $q = 3.0$}
\label{fig:qLargeConfig}
\end{figure}

\begin{figure}[h!]
\centering
\hfill
\subfigure[]
{\includegraphics[width=0.4\textwidth]{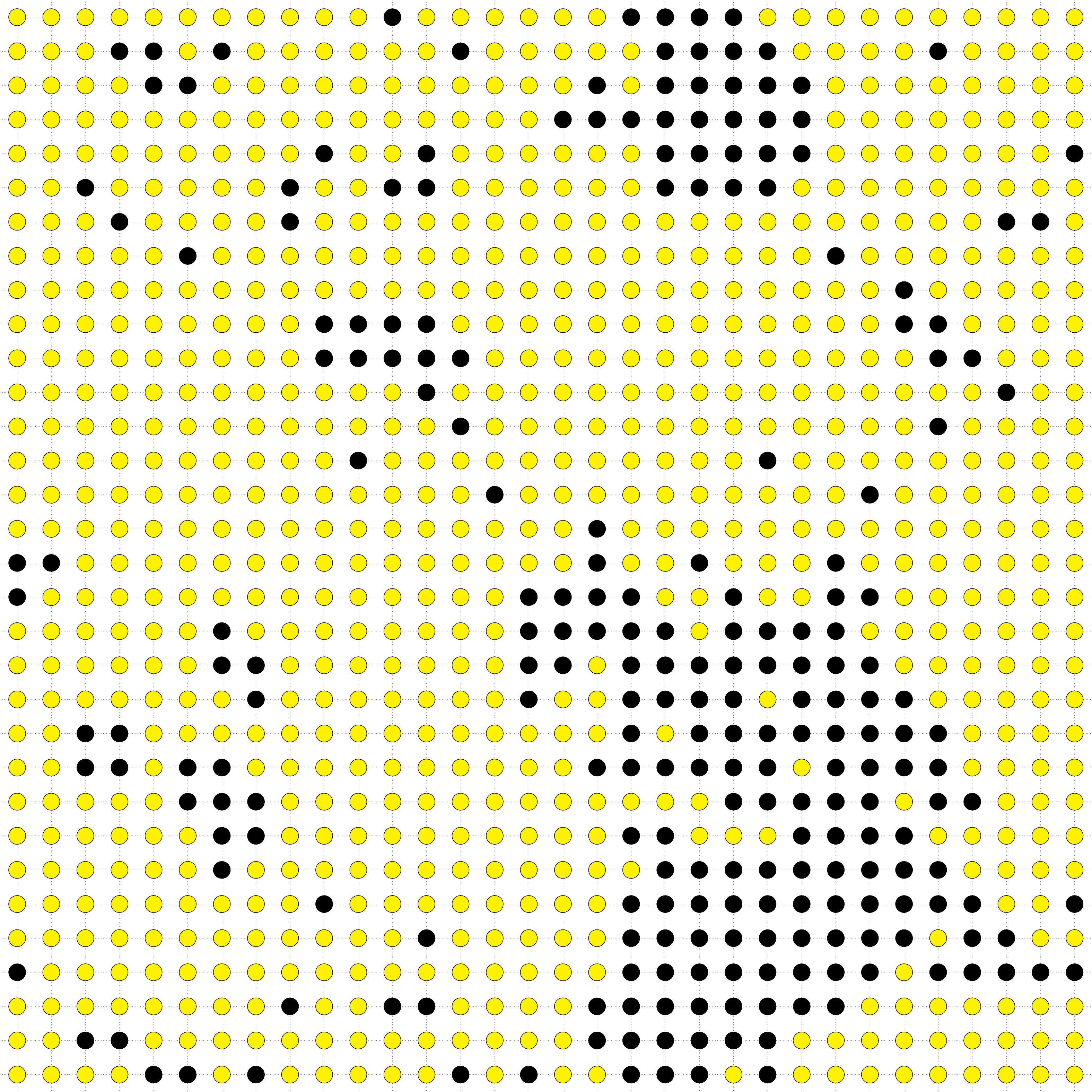}}
\hfill
\subfigure[]
{\includegraphics[width=0.4\textwidth]{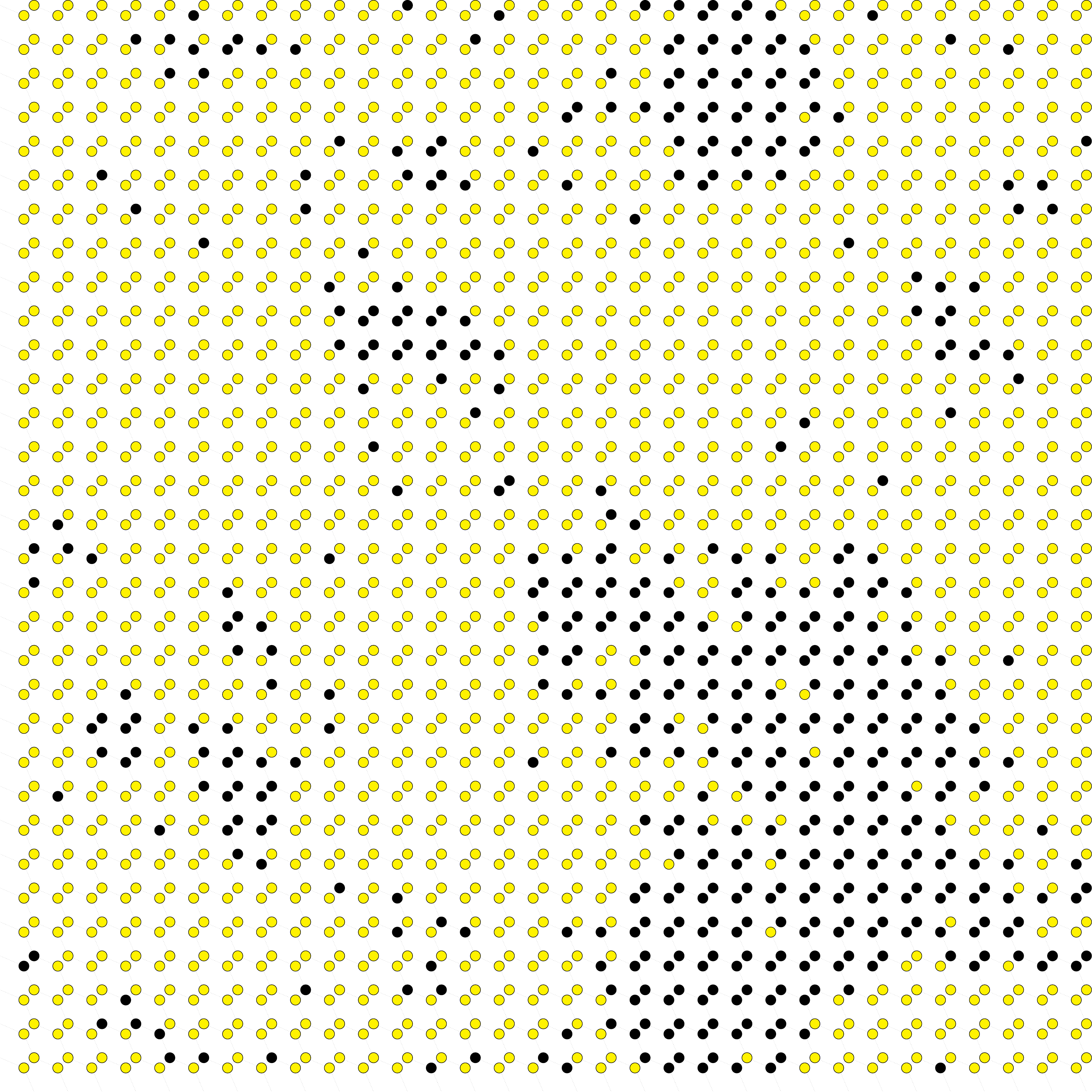}}
\hfill
\caption{$J = 0.6585$, $q = 0.6585$}
\label{fig:qBisectConfig}
\end{figure}

\begin{figure}[h!]
\centering
\hfill
\subfigure[]
{\includegraphics[width=0.4\textwidth]{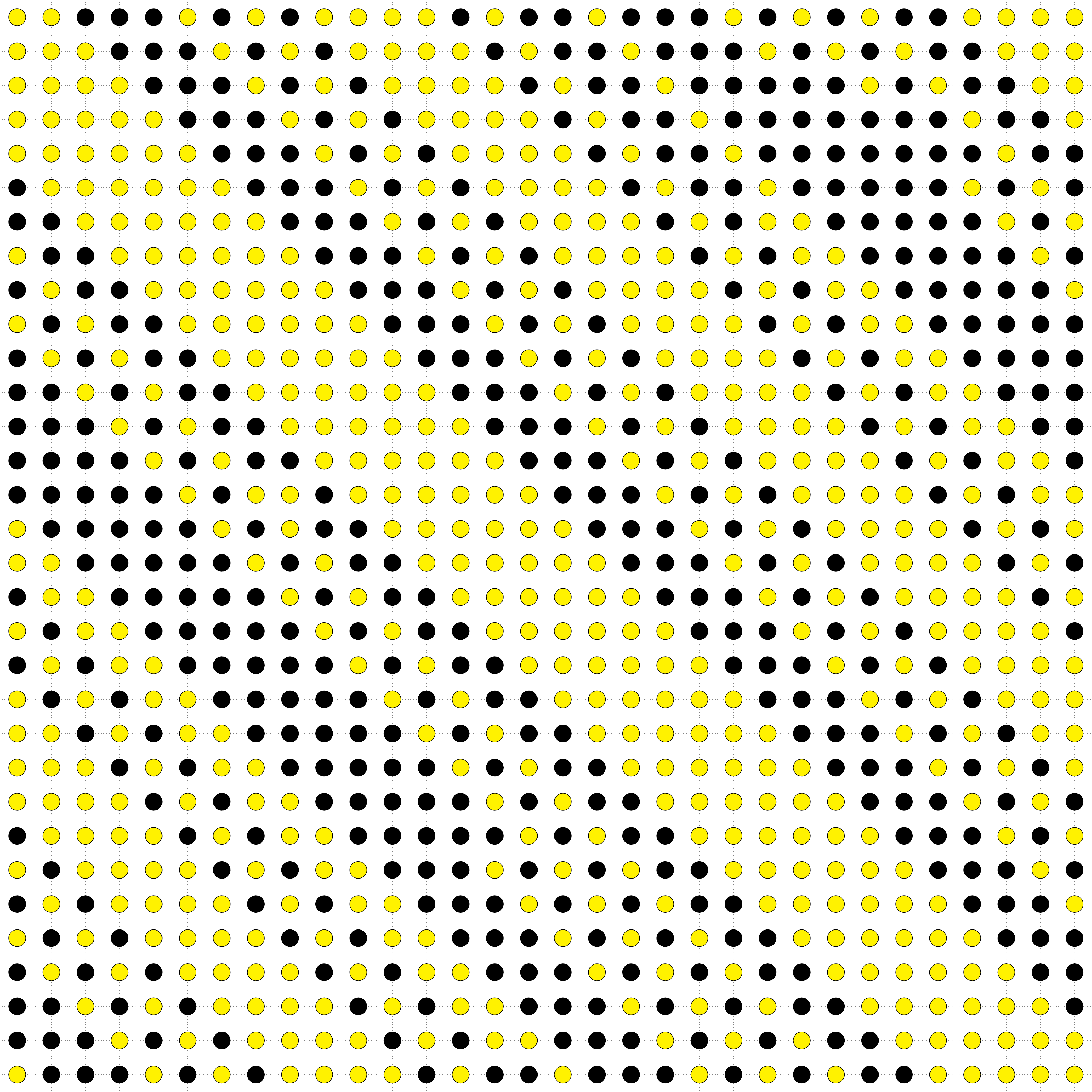}}
\hfill
\subfigure[]
{\includegraphics[width=0.4\textwidth]{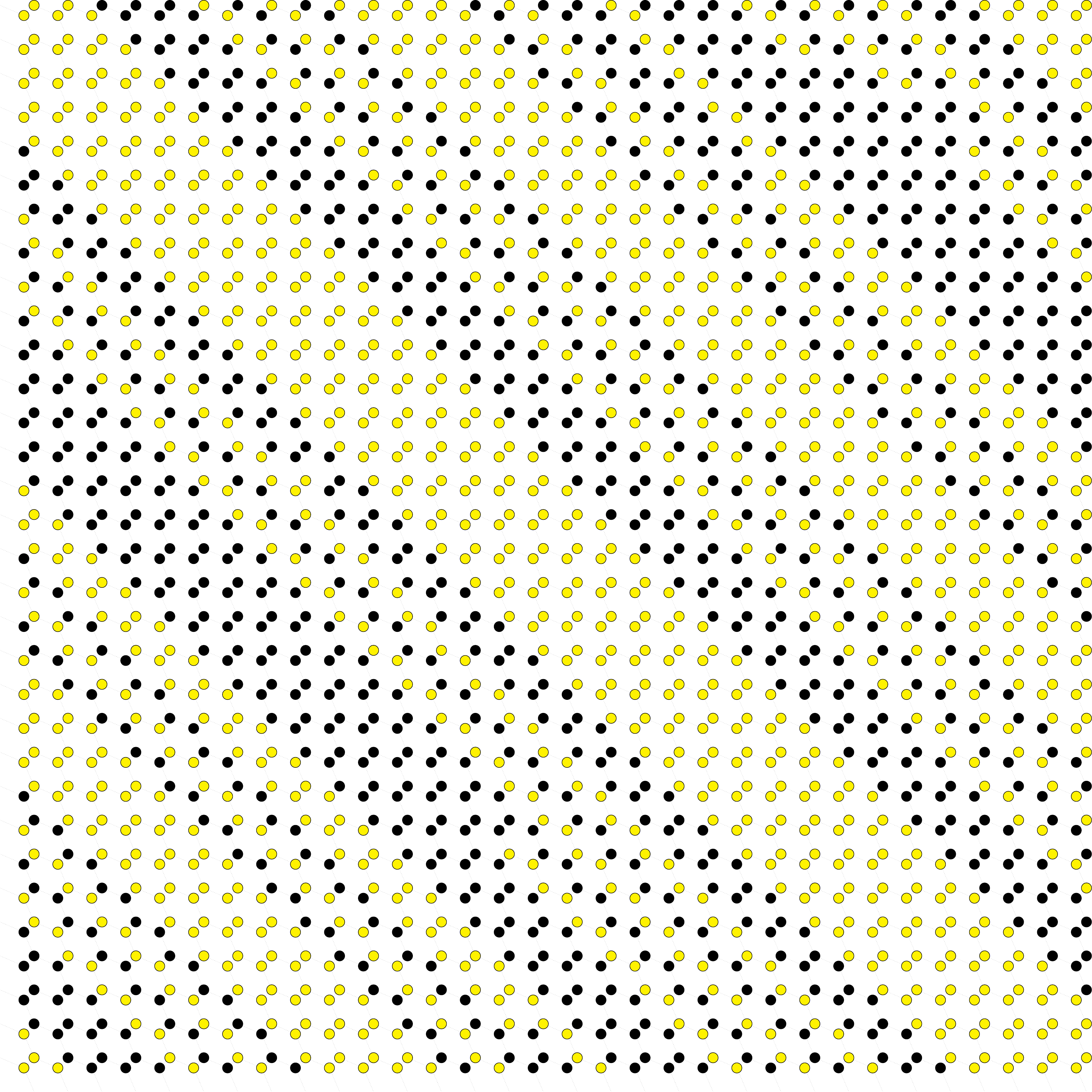}}
\hfill
\caption{$J = 2.0$, $q = 0.03$}
\label{fig:qSmallConfig}
\end{figure}

Figures~\ref{fig:qLargeConfig}, \ref{fig:qBisectConfig} and \ref{fig:qSmallConfig} show samples from the measure $\pi$ and the corresponding samples from the  Gibbs measure $\pi_2$ on $\cX_{\H}$ for several pairs of values
of $J$ and $q$ near the critical curve.

\begin{figure}[h!]
\centering
\hfill
\subfigure[]
{\includegraphics[width=0.45\textwidth]{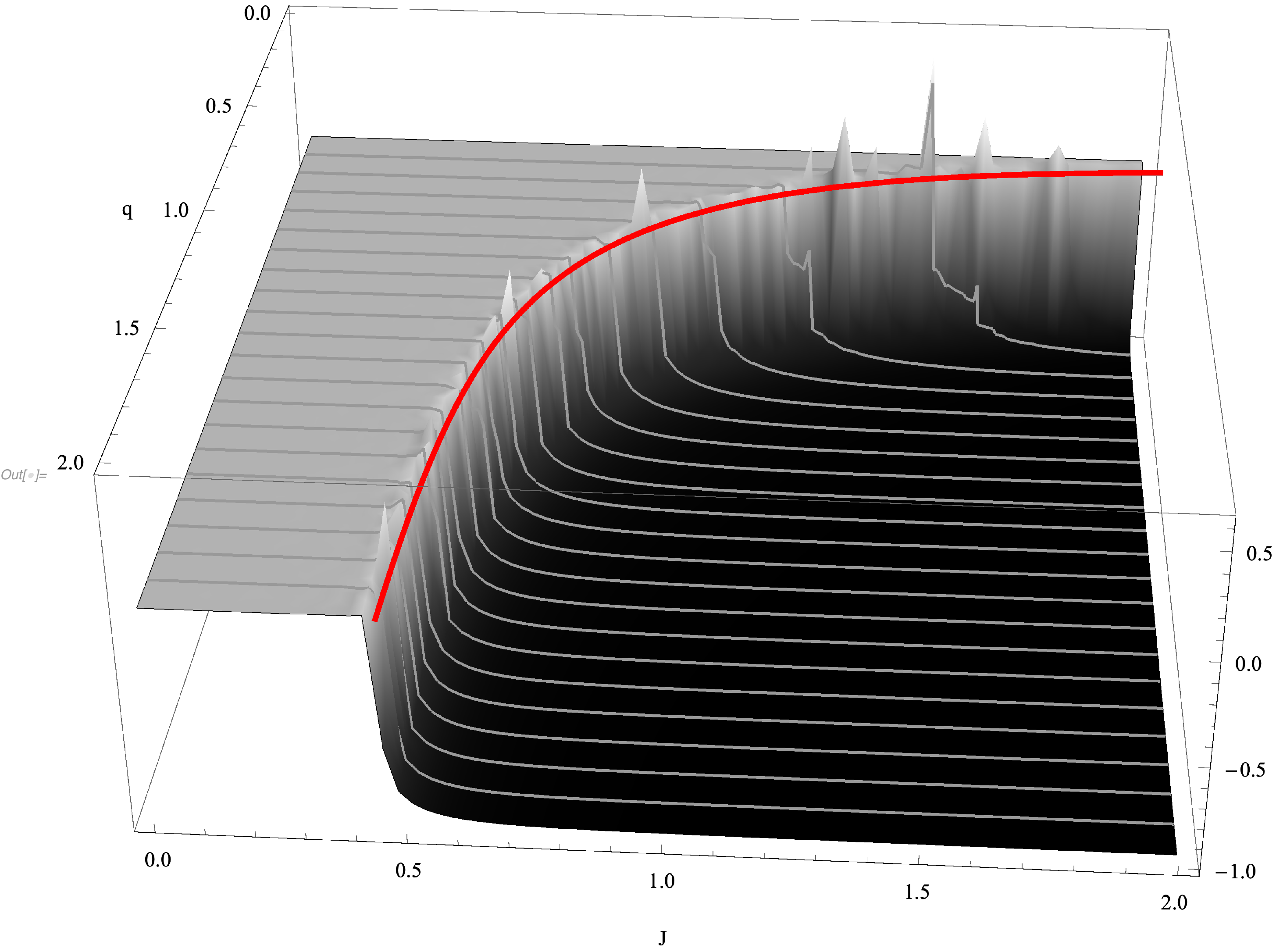}}
\hfill
\subfigure[]
{\includegraphics[width=0.45\textwidth]{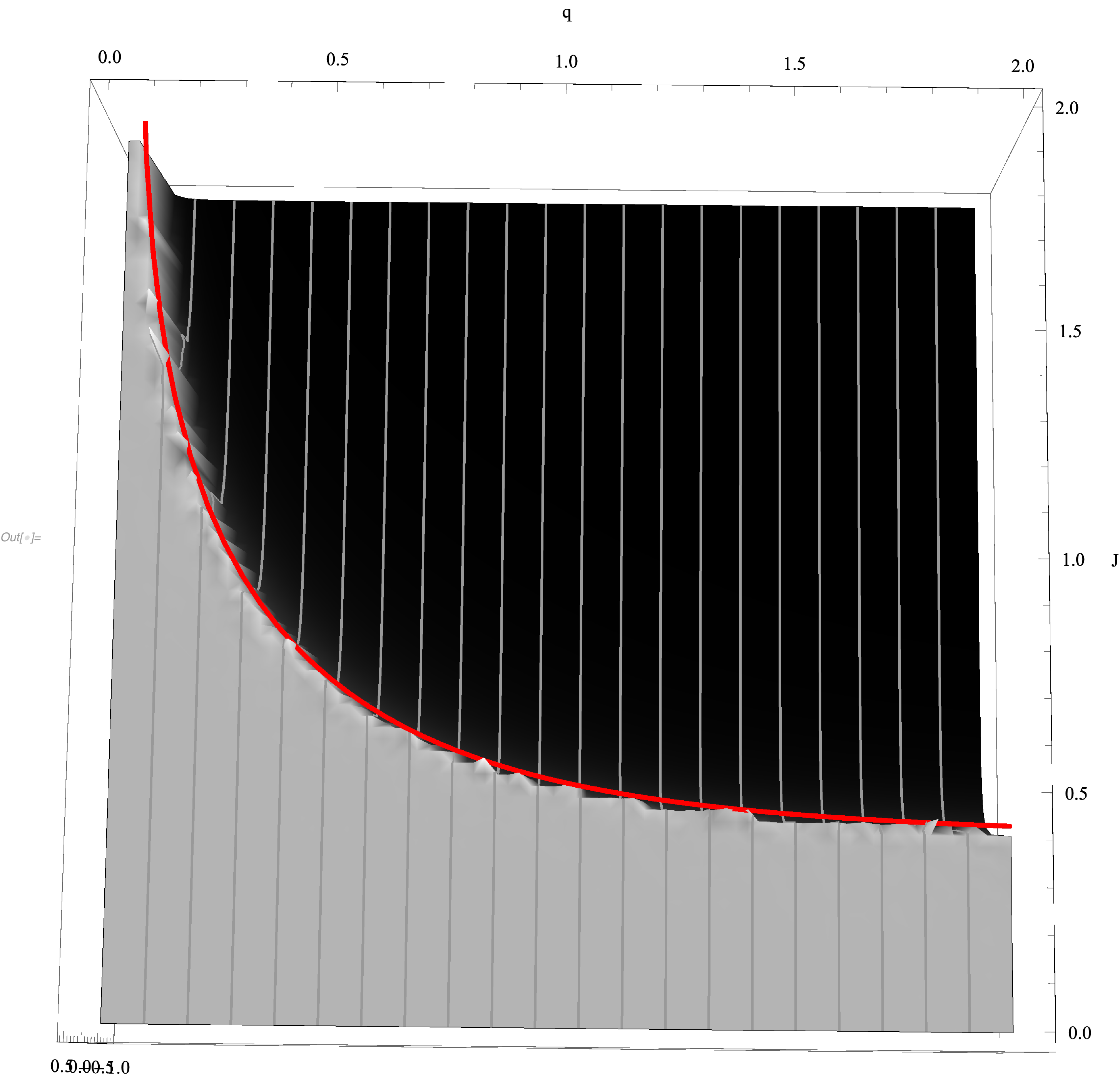}}
\hfill
\caption{The numerically determined critical curve}
\label{fig:numerical_critical_curve}
\end{figure}

Furthermore, numerical simulations show that
the responsivness of the ``shaken dynamics'' introduced in \cite{ADSST} to variations of the parameters is in very
good accordance with the theoretical results of Theorem~\ref{c1}. For instance, looking at
the average magnetization over a large number of iterations, it is possible to see that the
parameter space is clearly split into two regions corresponding to the ordered and disordered phase.
The numerically determined curve separating these two phases strongly agrees with the theoretical
one \eqref{eq:Jc_of_q} as shown in Figure~\ref{fig:numerical_critical_curve}.

{\bf Acknowledgments:}
We are grateful to Hugo Duminil-Copin for useful and interesting discussions.
B.S. acknowledges the MIUR Excellence Department Project awarded to
the Department of Mathematics, University of Rome Tor Vergata, CUP E83C18000100006. E.S. has been supported  by the PRIN 20155PAWZB ``Large Scale Random Structures". A.T. has been supported by Project FARE 2016 Grant R16TZYMEHN.
B.S. and E.S. thank
the support of the A*MIDEX project (n. ANR-11-IDEX-0001-02) funded by the  ``Investissements d'Avenir" French Government program, managed by the French National Research Agency (ANR).

\end{document}